\documentclass[prb,twocolumn]{revtex4-2}
\usepackage{geometry}
 \usepackage{natbib}
\usepackage{graphicx}
\usepackage{amsfonts}
\usepackage{amsmath}
\usepackage{bm}
\usepackage{url}
\usepackage{color}
\usepackage{xfrac}
\newcommand{\ev}{{\bm e}}
\newcommand{\gv}{{\bm g}}
\newcommand{\sv}{{\bm s}}
\newcommand{\xv}{{\bm x}}

\newcommand{\kv}{{\bm k}}

\newcommand{\Pm}{{\bm P}} 
\newcommand{\Mm}{{\bm M}}
\newcommand{\Rm}{{\bm R}}
\newcommand{\Em}{{\bm I}}
\newcommand{\Sm}{{\bm S}}

\definecolor{purple}{rgb}{0.5, 0.0, 0.8}
\definecolor{orange}{rgb}{0.9, 0.6, 0.0}

\begin{document}
\title{Quasicrystalline structure of the Hat monotile tilings} 

\author{Joshua E.~S.~Socolar}
\email[]{socolar@duke.edu}
\affiliation{Department of Physics, Duke University, Durham, NC, 27708}
\date{May 8, 2023}

\begin{abstract}
Tiling models can reveal unexpected ways in which local constraints give rise to exotic long-range spatial structure.  The recently discovered Hat monotile (and its mirror image) has been shown to be aperiodic~[Smith et al., arXiv:2303.10798 (2023)]; it can tile the plane with no holes or overlaps, but cannot do so periodically.  We show that the structure enforced by the local space-filling constraints is quasiperiodic with hexagonal (C6) rotational symmetry.  Although this symmetry is compatible with periodicity, the incommensurate ratio characterizing the quasiperiodicity stays locked to the golden mean as the tile parameters are continuously varied. We analyze a modification of the metatiles introduced by Smith et al.\ that yields a set of ``Key tiles'' that can be constructed as projections of a subset of six-dimensional hypercubic lattice points onto the two-dimensional tiling plane.  We analytically compute the diffraction pattern of a set of unit masses placed at the tiling vertices,  establishing the quasiperiodic nature of the tiling.  We point out several unusual features of the family of Key tilings and associated Hat tilings, including the tile rearrangements associated with the phason degree of freedom associated with incommensurate density waves, which exhibit novel features that may influence the elastic properties of a material in which atoms or larger particles spontaneously exhibit the symmetries of the Hat tiling. 
\end{abstract}

\maketitle 

\section{Introduction}
The recently discovered Hat tile has the remarkable property that congruent copies of it (together with its mirror image) can tile the plane but cannot do so in periodic manner.~\cite{smith2023aperiodic}  Smith et al.\ have shown  the Hat tilings can be viewed as decorations of tilings composed of ``metatiles'' that support a substitution symmetry.  Furthermore, they have shown that when the substitution operation is carried out ad infinitum, the tile shapes converge to ones with edge lengths related by the golden ratio.  The forcing of nonperiodic structure, the substitution symmetry, and the emergence of the golden ratio are all reminiscent of the well-known Penrose tiles~\cite{penrose1974}, which have served as a paradigmatic example of quasicrystalline structure, providing significant insights into the properties of physical quasicrystals.~\cite{levine1984,SteinOst,SteinDiV}   Smith et al. have shown that the Hat tile shape (or matching rules for the metatiles) forces some form of nonperiodic structure.  The purpose of the present paper is to provide a characterization of that structure, showing that these tilings are quasicrystalline but possess some novel symmetry properties.  

The signature of perfect quasicrystalline structure is a diffraction pattern consisting entirely of a dense set of Bragg peaks at wavevectors that are integer linear combinations of a set of incommensurate basis vectors.  In the Penrose tiling case, the natural choice of basis vectors is a 5-fold symmetric star, $\kv_n = k_0 \left(\cos(2\pi n/5),\sin(2\pi n/5)\right)$, for which $\kv_{n-1} + \kv_{n+1} = \phi^{-1}\kv_n$, where $\phi = (1+\sqrt{5})/2$ is the golden ratio.  One way to understand and calculate the diffraction pattern is based on viewing the tiling as a projection of a subset of points of 5D hypercubic lattice onto a 2D plane oriented such that the basis vectors of the hypercubic lattice project onto the five vectors that are edges of the rhombic Penrose tiles.~\cite{deBruijn1981,kramer1984}   A crucial feature of the construction is that the subset of 5D lattice points uniformly fills a bounded region (a ``window'') when projected instead onto the 3D space orthogonal to the tiling plane.  The diffraction pattern for a set of unit masses placed at the vertices of the Penrose rhombus tiling consists of Bragg peaks located at the projection of the 5D reciprocal space lattice onto the physical reciprocal space, where the peak intensities are determined by the Fourier transform of the window.~\cite{elser1985B,elser1986,jaric1986}

We show here that the Hat tiling can be constructed in an analogous manner.  It is a projection onto the tiling plane of a subset of 6D hypercubic lattice points that is bounded in a 4D subspace, though unlike the Penrose case, the window is not orthogonal to the tiling plane.  More conveniently, the Hat tiling can be viewed as a decoration of a set of four tiles that display the essential structure of the tiling.  Smith et al.\ made extensive use of four ``metatiles'' in proving the that Hat can only tile the plane in a nonperiodic pattern.  Here we modify their definition of the metatiles to form a set of ``Key tiles,'' so named because their vertices are the Key vertices identified by Smith et al.~\cite{smith2023aperiodic}  We show that the tilings formed by Key tiles have a simple structure in the 6D space.  

The natural class of Key tiles is larger than the set associated with Hat tiles.  From the physics perspective, all of these Key tilings have similar properties, and the special subset that permits a Hat decoration does not show any unique features other than allowing for a decomposition into a tiling by a Hat and its mirror image.  For a broader class of Key tilings, a decomposition into two distorted hat-like tiles, but the second is no longer the mirror image of the first.  For a broader class still, the decomposition algorithm produces overlapping tiles and one tile shape with a self-intersecting boundary.  This last class includes a case that we call the ``Golden Key,'' which plays a special role in the analysis of the entire class.

Smith et al.\ showed that the metatiles (or, equivalently, the Key tiles) in an infinite tiling can always be grouped into supertiles that are topologically equivalent to the originals and can be again grouped into still larger supertiles, ad infinitum.~\cite{smith2023aperiodic}  We refer to the operation of grouping tiles into larger tiles as ``inflation.''  Given an inflated tiling, the operation can be inverted to recreate a tiling of smaller tiles, a process we call ``deflation."~\cite{deBruijn1981,socolar1986}  All that is needed for the proof that the Hat tiling is nonperiodic is a proof that inflation operation is unique and can be iteratively performed ad infinitum on any infinite, space filling tiling.  The Golden Key tiles are the unique choices that allow for infinitely iterated deflation.  They correspond to the limiting shape produced by infinite inflations of any initial set of Key tiles; i.e., the Key tile set corresponding to the limiting metatile shapes identified by Smith et al.~\cite{smith2023aperiodic}  For generic Key tile sets, the tile shapes are not stable under deflation, and repeated iteration eventually produces tiles with perimeters that form figure-8 curves rather than simple polygonal regions.

This paper is organized as follows.  Section~\ref{keytiles} presents the lifting of Key tiling vertices onto a six-dimensional (6D) hypercubic lattice and defines the inflation and deflation operations in the 6D space.  This allows for a determination of the Golden Key tile shapes.  Section~\ref{hats} shows how a subclass of Key tilings can be decomposed into Hat tiles.  Section~\ref{diffraction} presents strong evidence (though not a rigorous proof) for the precise structure of the window that defines the relevant subset of hypercubic lattice points, then shows how this leads to the computation of the diffraction pattern for a set of unit masses located at the vertices of a Key tiling, which establishes the quasicrystalline nature of the tilings.  Finally, Section~\ref{remarks} contains remarks on unusual features of the Key tilings and Hat tilings and an illustration of the tile rearrangements associated with  infinitesimal phason shifts. 
 
\section{Key tilings as projections from a 6D hypercubic lattice} \label{keytiles}
\begin{figure}
\includegraphics[width=\columnwidth]{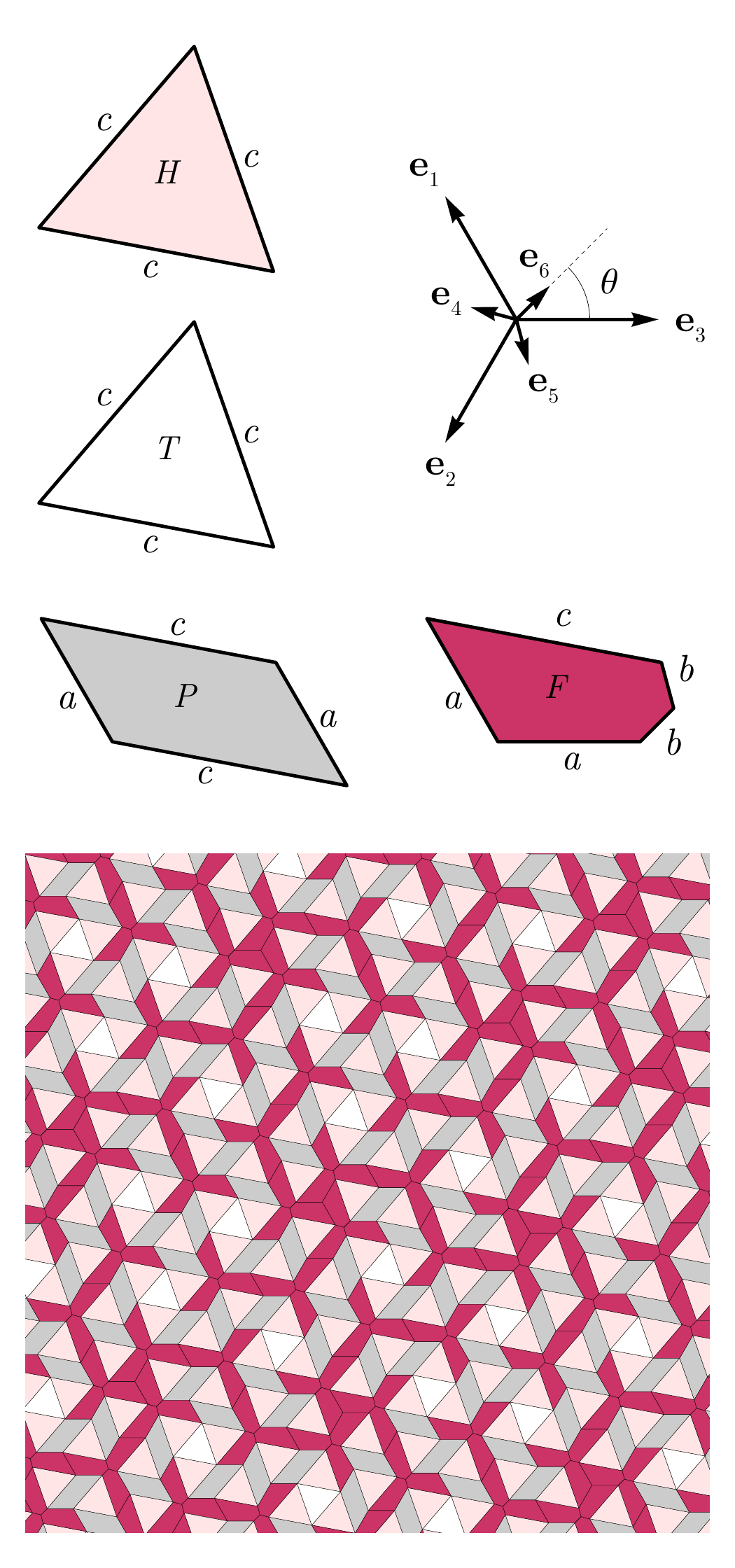}
\caption{Top: A generic set of Key tiles and the star vectors corresponding to the $a$ and $b$ edges with $\theta = \pi/4$.  The angles between adjacent $a$ edges and between adjacent $b$ edges are $120^{\circ}$.  Bottom: A portion of a tiling composed of this set of tiles.  Note that every vertex of the tiling lies on at least one $F$ tile.}
\label{fig:keytiles}
\end{figure}

\begin{figure}
\includegraphics[width=0.5\columnwidth]{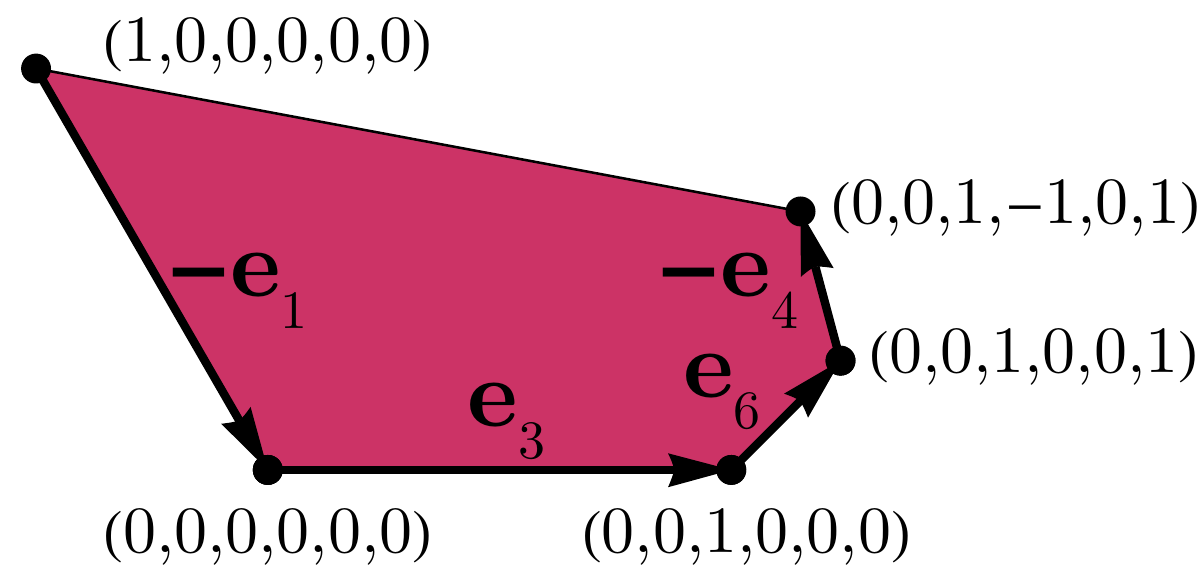}
\caption{Lifting an $F$ Key tile onto a 6D hypercubic lattice.  Displacements along each edge of length $a$ or $b$ correspond to the star vectors that lift into the basis vectors of the 6D lattice.}
\label{fig:lift}
\end{figure}

A set of Key tiles consists of the four shapes shown on the left in Fig.~\ref{fig:keytiles}.  
We use the same labels, $H$, $T$, $P$, and $F$, as those used by Smith et al.\ for their corresponding metatiles.~\cite{smith2023aperiodic}  The $a$ edges occur in three hexagonally symmetric orientations.  The $b$ edges also form a hexagonal set, but twisted by an angle $\theta$ from the set of $a$ edge orientations.  Edges of length $c$ connect vertices displaced by a vector that can be uniquely expressed as a sum of two vectors of length $a$ and two of length $b$.  We can thus lift the tiling into six dimensions by defining a set of six star vectors, ${\ev_n}$, in the tiling plane, arbitrarily choosing one tiling vertex as the origin, and indexing each vertex by the number of edges in each star direction that must be traversed from the origin to reach the vertex.  Fig.~\ref{fig:keytiles} shows the set of star vectors, and 
Figure~\ref{fig:lift} illustrates the construction of the lift to 6D.   
In this manner, each tiling vertex can be mapped to a lattice point in a 6D hypercubic lattice.  (The lift is uniquely determined up to a choice of the origin; i.e., there are no loops that could yield $\ev_1+\ev_2$ on one path and $-\ev_{3}$ on the other.)  The tiling vertices are projections of 6D lattice points onto a 2D plane oriented such that the basis vectors of the 6D lattice project onto the desired set of star vectors.  

\begin{figure}
\includegraphics[width=\columnwidth]{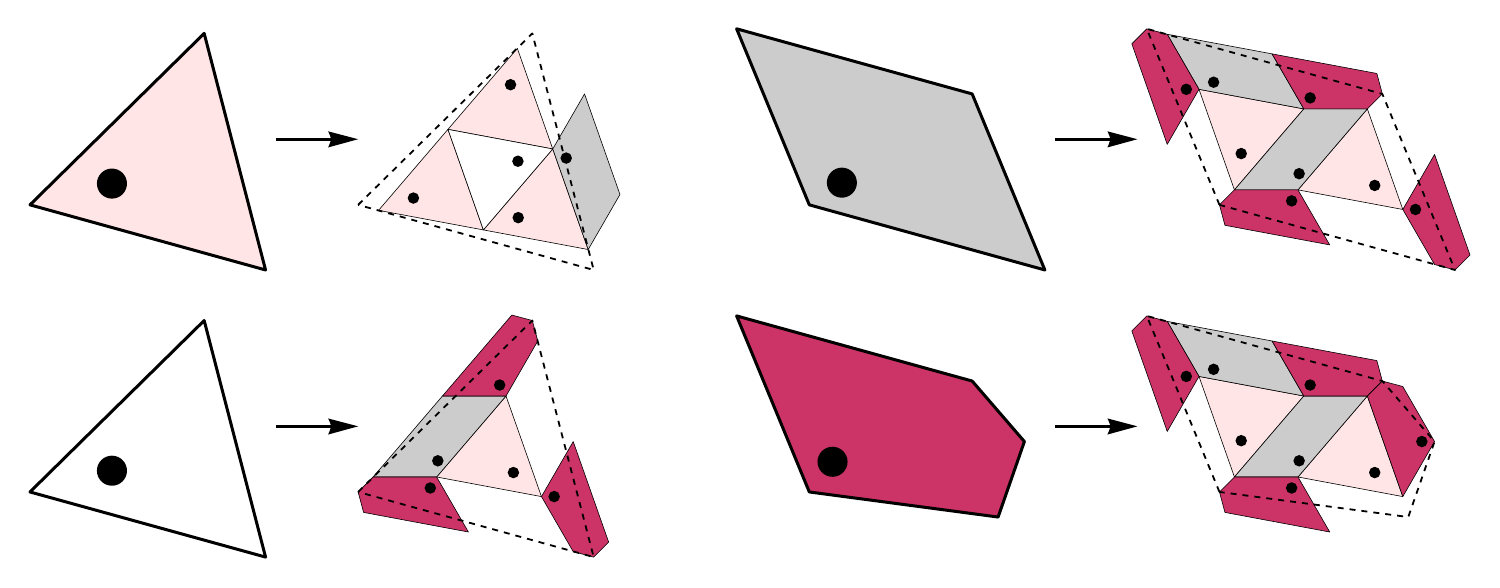}
\caption{The deflation operation for a generic set of Key tiles.  Black dots indicate the orientation of each tile for purposes of further inflation/deflation or placement of Hat decorations. Note that the deflated $P$ and $F$ tiles shapes (on the right) are not similar in the strict geometric sense to the originals, and the orientations of the $H$ and $T$ deflated tiles are slightly different from their parents.}
\label{fig:deflation}
\end{figure}

The deflation operation on Key tiles is depicted in Fig.~\ref{fig:deflation} and is equivalent to the substitution operation on the Smith metatiles~\cite{smith2023aperiodic}.  The smaller tiles are the originally defined Key tiles.  The larger ones represent one iteration of the inflation operation.  Note that the original and inflated $P$ and $F$ tiles have different shapes, and the original and inflated $H$ and $T$ tiles have different orientations.  As shown below, it is straightforward to derive the shapes of the inflated tiles given the original tiles.  Deriving the shapes of deflated tiles from an original set is more difficult, though in principle it can be done by inverting the relations obtained for the inflation operation.  In practice, to construct a tiling with a given set of tiles, one begins with those tiles and iterates the {\it inflation} operation multiple times to obtain the appropriate sets of tile shapes for each iteration of the inflation operation.  One can then begin with a high-level inflated tile and perform the deflation operation repeatedly, using the tile shapes that have already been determined, to construct a large patch of the tiling.

The substitution matrix for the numbers of tiles generated by deflation is 
\begin{equation}
\Mm = \left(\begin{array}{cccc}
3 & 1 & 2 & 2 \\
1 & 0 & 0 & 0 \\
1 & 1 & 2 & 2 \\
0 & 3 & 4 & 4 \end{array}\right),
\end{equation}
where each column represents the numbers of $H$, $T$, $P$, and $F$ tiles present in the deflation of each type of tile, with the columns listed in the same order.
The largest eigenvalue is $\lambda_1 = \phi^4 = 6.854…$, and the eigenvector associated with it gives the following relative frequencies $\rho_X$ of the four tile types in the infinite tiling:
\begin{equation} \label{eqn:ratios}
\rho_H: \rho_T: \rho_P:\rho_F = \phi^4 : 1 : 3\phi : 3\phi^2.
\end{equation}
Note that the second largest eigenvalue is $\lambda_2 = 2$, which is greater than unity.  For one-dimensional substitution tilings, a second eigenvalue greater than unity indicates strong fluctuations that give rise to continuous components of the diffraction spectrum.~\cite{bombieri1986}.  In two dimensions, the situation is more complicated.  We note that in the present case $\lambda_1$ is a Pisot number and $\lambda_2< \lambda_1^{1/2}$, which indicates that under repeated deflation of a patch of tiles, fluctuations in tile densities scale more slowly than the perimeter of patch.  For analysis of the implications of these features, we refer the reader to the literature on spectra of $d$-dimensional substitution tilings~\cite{godreche1992,sadun2011,schmieding2018}. For present purposes, we will provide strong numerical evidence below that the Key tilings exhibit pure point diffraction, as rigorously confirmed in Ref.~\cite{baake2023}.
 
We now consider the tile shapes generated by repeated inflation of the tiles.
Let $\xv$ be a 6D integer vector representing an edge in the original tiling, and let $\Rm$ be the matrix that acts on $\xv$ to rotate the corresponding edge by $\pi/3$ in the tiling plane.  Inspection of the set of star vectors as they are indexed in Fig.~\ref{fig:keytiles} immediately allows us to construct 
\begin{equation}
\Rm = \left( \begin{array}{cccccc}
0 & 0 & -1 & 0 & 0 & 0 \\
-1 & 0 & 0 & 0 & 0 & 0 \\
0 & -1 & 0 & 0 & 0 & 0 \\
0 & 0 & 0 & 0 & 0 & -1 \\
0 & 0 & 0 & -1 & 0 & 0 \\
0 & 0 & 0 & 0 & -1 & 0 
\end{array}\right).
\end{equation}

\begin{figure}
\includegraphics[width=\columnwidth]{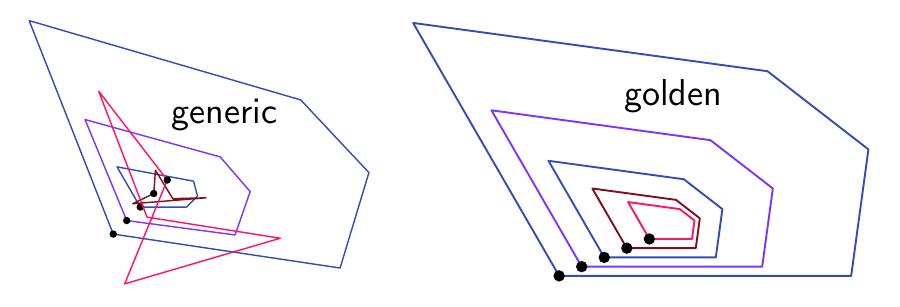}
\caption{Inflations and deflations of an $F$ tile.  Left: A generic case.  Inflation produces a sequence of larger tile shapes that converge to the Golden Key $F$ tile shape, while deflation at some point generates inverted or self-intersecting tiles.  Right: The Golden Key case.  Inflation and deflation preserve the shape and orientation of the tile at each step, so deflation can be carried out ad infinitum.  Tiles deflated $n$ times have been scaled by factors of $\phi^{n/2}$ in the generic case and $\phi^n$ in the golden case for visualization purposes.  Dots mark the corresponding vertices on the series of deflated tiles.}
\label{fig:deflatingShapes}
\end{figure}

Let $\ev^{(n)}_i$ be the 6D lift of the $i^{th}$ star vector associated with the $n^{th}$ inflation of the original Key tiles, where $\ev^{(0)}_i$ corresponds to the 6D lift of the corresponding star vector shown in Fig.~\ref{fig:keytiles}.
The inflation operation consists of grouping sets of edges in the original tiling together to form edges of the inflated tiles.  Inspection of the $F$ tile in Fig.~\ref{fig:deflation} reveals that the vector corresponding to the bottom $a$ edge of the inflated tile is
\begin{eqnarray}
\ev^{(n+1)}_3 & = & \ev^{(n)}_3 + \ev^{(n)}_6 + \Rm^5 \cdot \ev^{(n)}_3 +\ev^{(n)}_6 + \ev^{(n)}_3 \\
\  & =  & (2 \Em + \Rm^5) \cdot \ev^{(n)}_3 + 2\Em \cdot  \ev^{(n)}_6,
\end{eqnarray}
where $\Em$ is the $6\times 6$ identity matrix.
Similarly, we have
\begin{equation}
\ev^{(n+1)}_6  = \Rm \cdot \ev^{(n)}_3 + \Rm \cdot  \ev^{(n)}_6.
\end{equation}
Defining $\sv^{(n)}$ as the 12-component concatenation of $\ev^{(n)}_3$ and $\ev^{(n)}_6$, we have
\begin{equation}
\sv^{(n+1)} = \Sm\cdot\sv^{(n)}; \quad \Sm = \left(\begin{array}{cc} (2\Em + \Rm^5) & 2\Em  \\  \Rm & \Rm  \end{array}\right).
\end{equation}
$\Sm$ has four degenerate eigenvalues $\phi^2$, four pure imaginary eigenvalues $\pm i$, and four degenerate eigenvalues $\phi^{-2}$.  We begin with the 6D lifts of star vectors $\ev_3$ and $\ev_6$; i.e., $\sv^{(0)}
= (0,0,1,0,0,0,0,0,0,0,0,1)$.  Decomposing $\sv^{(0)}$ into a sum of eigenvectors of $\Sm$, we find
\begin{equation}
\lim_{n\rightarrow \infty} \sv^{(n)} = \phi^{2n} \sv^{\ast}
\end{equation}
where $\sv^{\ast}$ is a vector in the subspace spanned by the eigenvectors with eigenvalue $\phi^2$:
\begin{align}
\sv^{\ast} & = \frac{1}{3\sqrt{5}}(  2+\phi\,,\,\phi^{-1}\,,\,-\phi^3\,,\, 2\,,\, 2\,,\, -4\,, \nonumber \\
 \  &  \quad\quad\quad\quad -1\,,\, 2\,,\, -1\,,\,  \phi-3\,,\, \phi\,,\,  -\phi^{-3} \,) \\
   \ & \equiv  (\ev^{\ast}_3,  \ev^{\ast}_6).
\label{eqn:evast}
\end{align}
The vectors $\ev^{\ast}_3$ and $\ev^{\ast}_6$ define a 2D plane in the 6D space that is invariant under inflation, and the Golden Key tiles lie in this plane.  Thus the 6D vectors $\ev^{\ast}_i$ can be taken to be the star vectors $\ev_i$ of the Golden Key tiling, and the angle $\theta$ for this case can be calculated from them:
\begin{align}
\theta^{\ast} & =  \cos^{-1} \left(\frac{\ev^{\ast}_3\cdot\ev^{\ast}_6}{||\ev^{\ast}_3||\,||\ev^{\ast}_6||}\right) =  \cos^{-1} \left(\frac{\sqrt{2}}{4\phi^2}\right).
\end{align}
The Golden Key tile shapes are unique in that they can be deflated infinitely many times.  For any other choice of Key tile star vectors, the 6D star vectors must be projected onto a 2D plane to create a planar tiling, and the projection of the 6D star vectors onto the eigenvectors of $\Sm^{-1}$ that grow under deflation will at some iteration produce self-intersecting 2D tile shapes, as illustrated in Fig.~\ref{fig:deflatingShapes}.

\begin{figure}
\includegraphics[width=\columnwidth]{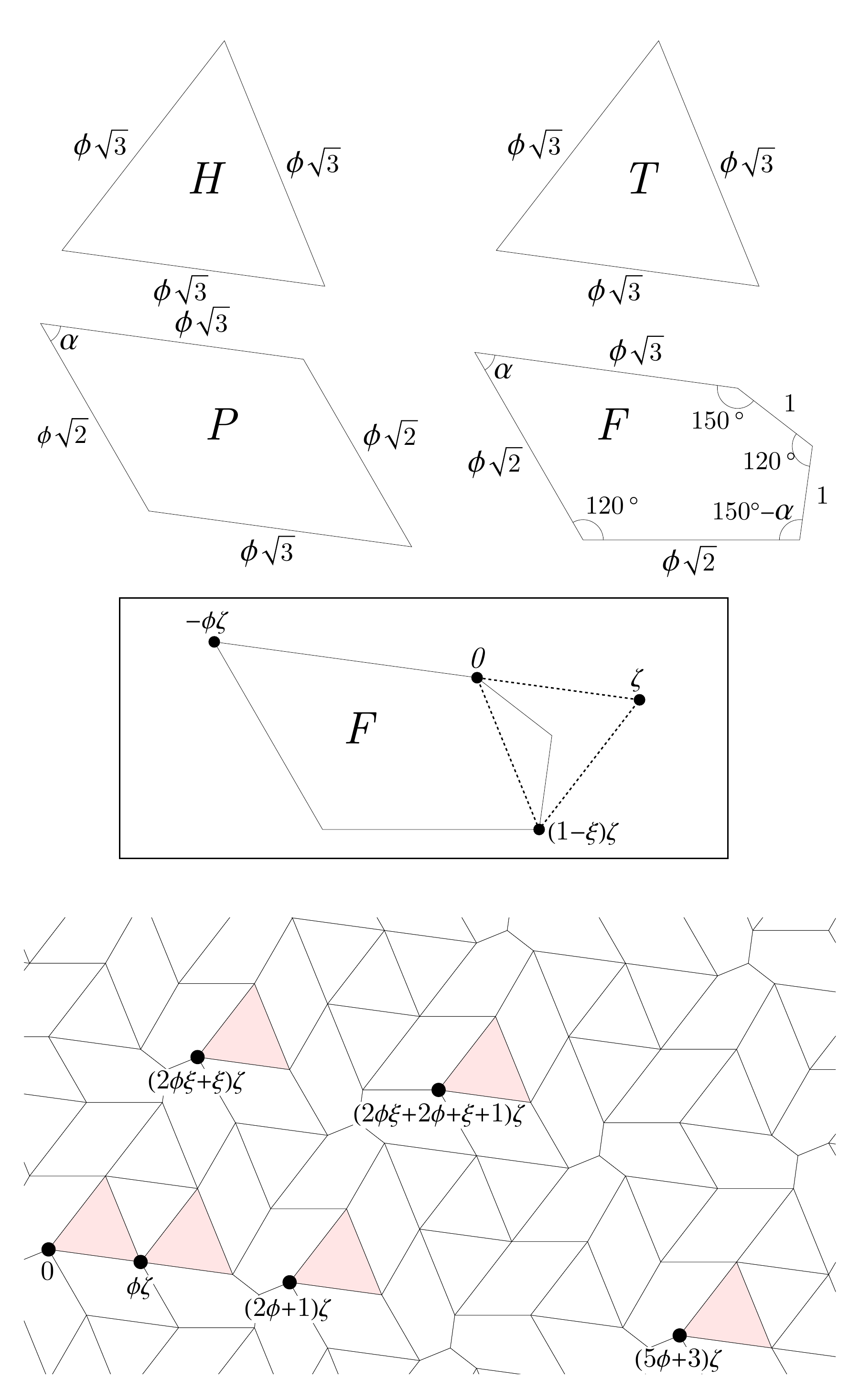}
\caption{The set of Golden Key tiles with edge lengths and angles labeled.  $\phi = (1+\sqrt{5})/2$ is the golden mean, and $\alpha = \theta^{\ast}-\pi/6$.  The boxed tile and tiling patch illustrate the geometry of the set of return vectors.  (See text.)  The shaded tiles in the tiling patch are all $H$ tiles with the same orientation.}
\label{fig:golden}
\end{figure}

Figure~\ref{fig:golden} shows the dimensions of the Golden Key tiles.  A key feature of the analysis by Baake et.~al~\cite{baake2023} is the determination of the set of return vectors of the tiling; i.e., the displacements required to bring two tiles of the same type and orientation into coincidence. The boxed $F$ tile shows that three points on the $F$ tile correspond to certain complex numbers, where $\xi = e^{i\pi/3}$ and $\zeta$ is a constant setting the scale and orientation of the tiling.  The tiling patch illustrates the fact that the return vectors of the Golden Key tiling are members of $\zeta\mathbb{Z}[\phi,\xi]$, where $\mathbb{Z}[\phi,\xi]$ is the set of all integer-coefficient linear combinations of $1$, $\phi$, $\xi$, and $\phi\xi$, consistent with Ref.~\cite{baake2023}.

\section{Hat tilings}\label{hats} 
\begin{figure}
\begin{center}
\includegraphics[width=0.9\columnwidth]{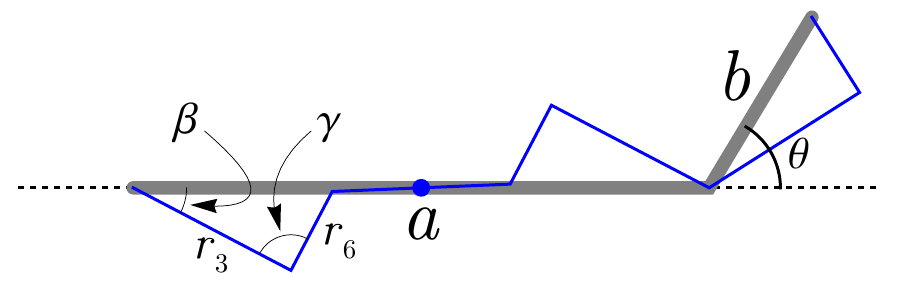}
\end{center}
\caption{The relation between Hat edges (thin lines) and $F$ Key tile edges (thick lines).  The Key tile parameters match
Fig.~\ref{fig:keytiles}.  The angles between consecutive $r_3$ edges and between consecutive $r_6$ edges are both $2\pi/3$.}
\label{fig:FedgesHat}
\end{figure}

\begin{figure}
\begin{center}
\includegraphics[width=1.0\columnwidth]{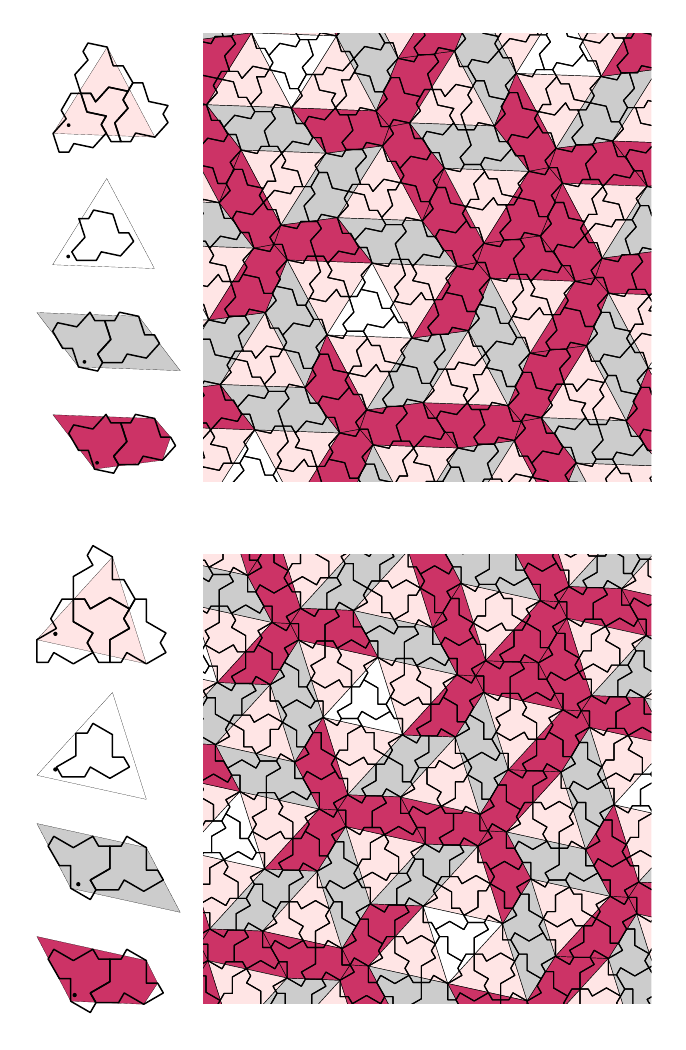} 
\end{center}
\caption{Two Hat tilings.  Top: A generic case with $\gamma > \pi/2$.  Note that the Hat at the center of the $H$ tile (top left) is {\em not} isomorphic to the other hats.  Bottom: a case with $\gamma = \pi/2$.  Here the Hat at the center of the $H$ tile is a mirror image of the primary hat, so the tiling consists of a single tile shape in the class discovered by Smith et al.}
\label{fig:HatTilings}
\end{figure}

We now consider the sets of Key tiles that permit the Hat decoration.
Figure~\ref{fig:FedgesHat} shows the relation between Hat edges and the $a$ and $b$ edges of an $F$ Key tile.
The tiles specified as Tile[$r_6,r_3$] in Ref.~\cite{smith2023aperiodic} correspond to the special case $\gamma = \pi/2$.
The angle $\gamma$ and the lengths $r_3$ and $r_6$ define the Hat shape, as shown in Fig.~\ref{fig:HatTilings}.
The associated Key tiles have parameters (defined in Fig.~\ref{fig:keytiles})
\begin{align}
a & = \sqrt{r_3^2 + 3 r_6^2-2\sqrt{3}r_3 r_6 \cos\left(\tfrac{\pi}{6} +\gamma\right)}, \\
b & = \sqrt{r_3^2 + r_6^2 - 2 r_3 r_6 \cos\gamma}, \\
\theta & = \frac{\pi}{3} - \beta + \chi, 
\end{align}
where 
\begin{align} 
\beta  & =   \cos^{-1}\left(\frac{r_3 -\sqrt{3} r_6 \cos\left(\frac{\pi}{6} + \gamma\right) }{\sqrt{r_3^2 + 3 r_6^2 - 2\sqrt{3}\, r_3 r_6 \cos\left(\frac{\pi}{6} + \gamma\right) }}\right), \label{eqn:beta} \\
\chi & = \cos^{-1}\left(\frac{r_3 - r_6 \cos\gamma }{\sqrt{r_3^2 +r_6^2 - 2 r_3 r_6 \cos\gamma }}\right). \label{eqn:chi}
\end{align}

Figure~\ref{fig:HatTilings} shows two Key tilings with Hat decorations.  Note that the ``reflected'' Hat that appears in each $H$ tile is a mirror image of the primary Hat if and only if $\gamma = \pi/2$.  Note also that not all values of $b/a$ and $\theta$ for the Key tiles can be obtained from choices of $r_3$, $r_6$, and $\gamma$.  In particular, the Golden Key parameters cannot be produced from Hat decorations.  A complete solution for the set of parameters of Key shapes that admit Hat tilings is beyond the scope of this paper. 

It appears most natural to conceive of the Hat tiles as a decoration of the Key tiles.  Many other decorations are possible, of course, and it is not clear whether the Hat tilings have any special properties other than the fact that their shapes alone force a nonperiodic tiling and that a one-parameter family of them become monotiles by reflection symmetry.  We note also that the Hat tilings themselves can be lifted onto a 6D hypercubic lattice, where now the edges of the lattice project onto the six edge vectors that form the hats.  The vertices of the associated Key tiling are a subset of the Hat vertices.  We will see below that the lift of the Hat tiling vertices presents a substantially more complex structure than the lift of the Key tiling vertices. 

\section{6D structure and diffraction}\label{diffraction}
\begin{figure}
\includegraphics[width=0.9\columnwidth]{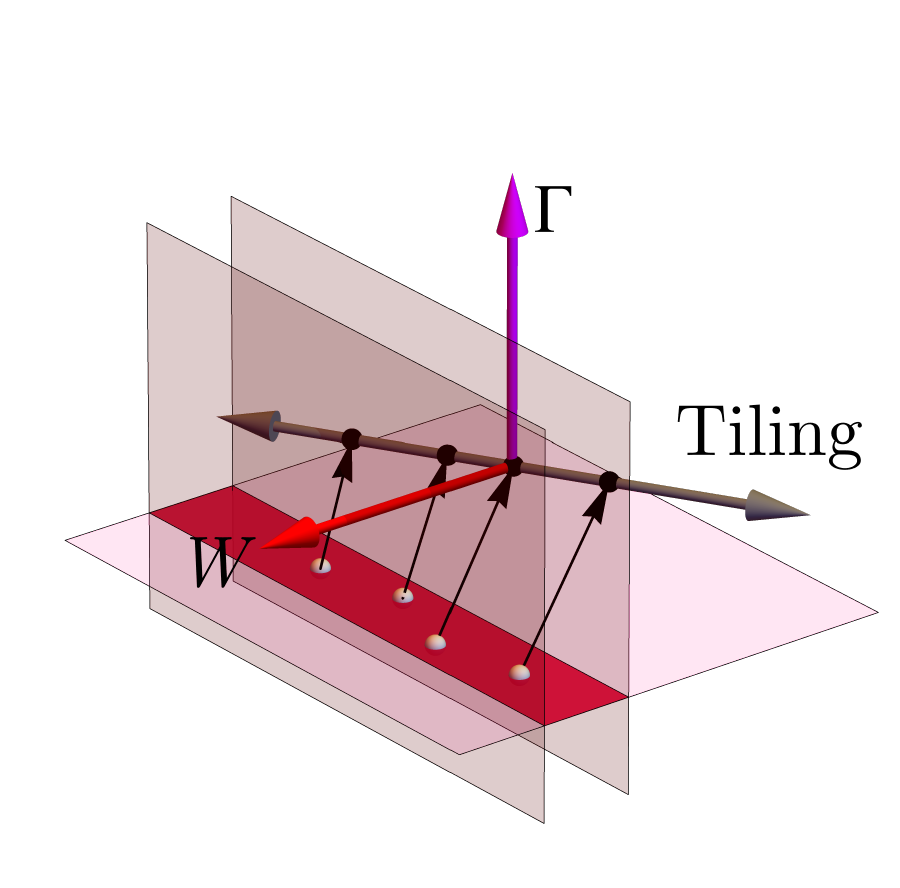}
\caption{Schematic illustration of the projection construction of the tiling vertices.  Each vector shown here represents a 2D space.  Small spheres represent points that have the same projection onto $\Gamma$ (horizontal plane) and lie within a certain window in $W$ (between the two vertical planes).  This set of points is projected onto the tiling space to form the tile vertices.  While $W$ and the tiling space are both orthogonal to $\Gamma$, they are {\em not} orthogonal to each other.}
\label{fig:Proj}
\end{figure}

To characterize the type of order displayed by the Key tilings, we compute their diffraction patterns.  The computation presented here is based on strong numerical evidence for the shape of the 4D acceptance window that determines which 6D hypercubic lattice points project to the vertices of the tiling.  A rigorous derivation of this window shape has not yet been completed.  The logic of the calculation is as follows:

\noindent $\bullet$ We first show that the projection of the 6D tiling vertices onto the plane spanned by  $(1,1,1,0,0,0)$ and $(0,0,0,1,1,1)$ consists of just four points.  We refer to this 2D subspace as $\Gamma$.  This is schematically illustrated in Fig.~\ref{fig:Proj} as points lying in a horizontal plane normal to the $\Gamma$ direction, all of which project into one point in $\Gamma$ in the illustration.

\noindent $\bullet$ We then show that the 6D tiling vertices project into a certain compact window in a 2D subspace $W$ that is orthogonal to $\Gamma$.  The window is illustrated in Fig.~\ref{fig:Proj} as the segment of the line in the $W$ direction that lies between the two vertical planes normal to $W$.  (The figure also illustrates the projection of lattice points onto the tiling plane, which is orthogonal to $\Gamma$ but but not to $W$.)  The window in $W$ consists of four equilateral triangles, each corresponding to a different point in $\Gamma$, and each of which is filled densely and uniformly with the same density.   Each triangle is the 2D analogue of the segment in Fig.~\ref{fig:Proj} corresponding to a particular point in $\Gamma$.  (Reference~\cite{baake2023} presents an alternative approach to the description of the tiling as a projection of a higher dimensional lattice.)

\noindent $\bullet$ We then carry out a slight modification of a standard technique for computing the diffraction pattern of projected (quasicrystal) structures.~\cite{elser1986,jaric1986}  Each 6D reciprocal space lattice vector $\kv$ can be decomposed into a sum $\kv_t + \kv_w + \kv_g$, where the subscripts indicate vectors lying in the reciprocal space of the tiling plane, of $W$, and of $\Gamma$, respectively.   This allows us to express the Fourier transform of a set of unit masses placed at the tiling vertices in terms of a sum of the Fourier transforms of the four triangular windows with relative phase factors determined by the corresponding positions in $\Gamma$.
We note that more recently developed techniques for determining the dynamical spectrum of a substitution tiling have been applied to the present case by Baake et al.~\cite{baake2023} to yield a rigorous proof that the spectrum consists entirely of Bragg peaks.  That approach leads to a projection scheme involving a 4D lattice, and we note that the 6D construction adopted in the present work could in principle be reduced to 4D by writing $\ev_3 = -\ev_1+\ev_2$ and $\ev_6 = -\ev_4+\ev_5$, effectively collapsing the $\Gamma$ subspace.

The diffraction pattern consists of delta-function (Bragg) peaks at the dense set of points $\kv_t$ that are projections of the hypercubic reciprocal space lattice, each with an amplitude determined by the corresponding $\kv_w$ and $\kv_g$.  For a given 6D lattice vector $\kv$, the components $\kv_t$, $\kv_w$, and $\kv_g$ are determined by $(\Pm^{-1})^T \kv$, where $\Pm$ is the matrix that converts hypercubic lattice point coordinates in the standard orthonormal basis into coordinates in a basis consisting of pairs of unit vectors spanning each of the three subspaces: 
\begin{equation}
(\xv_t,\xv_w,\xv_g) = \Pm\cdot\xv,
\end{equation}  
where
\begin{widetext}
\begin{equation}
\Pm = \left(\begin{array}{cccccc}
a \cos(\tfrac{2\pi}{3}) & a \cos(\tfrac{4\pi}{3}) & a \cos(\tfrac{6\pi}{3}) & b\cos(\theta +\tfrac{2\pi}{3}) & b\cos(\theta +\tfrac{4\pi}{3}) & b\cos(\theta + \tfrac{6\pi}{3}) \\
a \sin(\tfrac{2\pi}{3}) & a \sin(\tfrac{4\pi}{3}) & a \sin(\tfrac{6\pi}{3}) & b\sin(\theta +\tfrac{2\pi}{3}) & b\sin(\theta +\tfrac{4\pi}{3}) & b\sin(\theta +\tfrac{6\pi}{3}) \\

\phi \nu & (\phi-3)\nu & -\phi^{-3}\nu & -2\nu & \nu & \nu  \\
-\phi^{-2}\mu & -\phi^{-1}\mu & \mu & 0 & \mu &- \mu \\

1/\sqrt{3} & 1/\sqrt{3} & 1/\sqrt{3} & 0 & 0 & 0 \\ 
0 & 0 & 0 & 1/\sqrt{3} & 1/\sqrt{3} & 1/\sqrt{3} .
\end{array}\right)
\end{equation}
\end{widetext}
Here $a$, $b$, and $\theta$ are the edge lengths and angle defined in Fig.~\ref{fig:keytiles}, where $a$ and $b$ have been scaled such that the row is normalized to unity, and $\nu$ and $\mu$ are factors that normalize their respective rows to unity. The first two rows of $\Pm$ are the $x$ and $y$ coordinates of the projections of the 6D basis vectors onto the tiling plane.  The fifth and sixth rows are unit vectors in the $\Gamma$ subspace.  The third and fourth rows are orthogonal coordinates in the $W$ subspace, each being orthogonal to the $\Gamma$ subspace and to both $\ev_3^{\ast}$ and $\ev_6^{\ast}$ as defined in Eq.~(\ref{eqn:evast}).  This ensures that the $W$ subspace is orthogonal to the the 6D vectors representing the edges of infinitely inflated tiles. 

\begin{figure}
\includegraphics[width=\columnwidth]{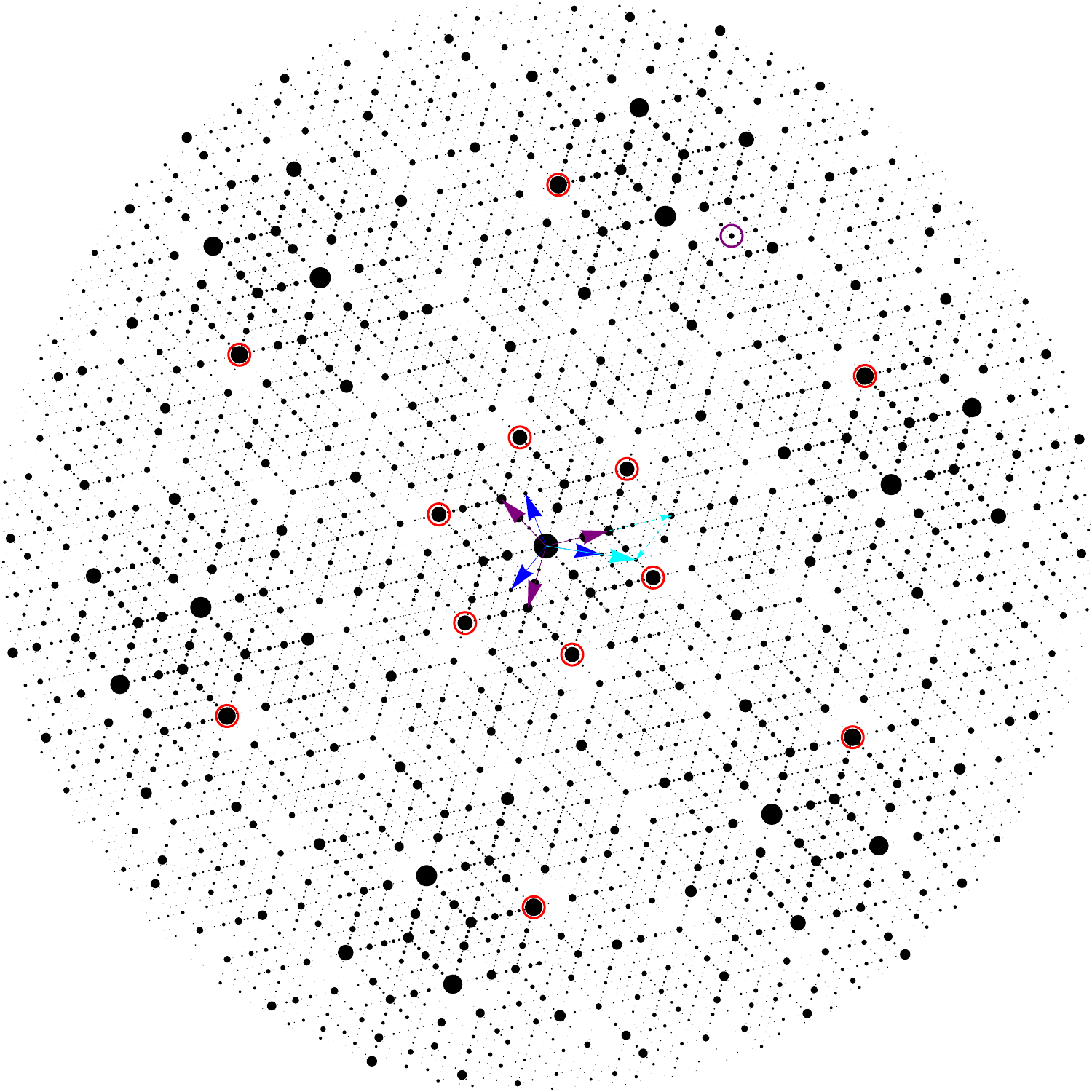}
\caption{The diffraction pattern of unit masses placed at the vertices of the tiling of Fig.~\ref{fig:keytiles}.  Disk area is proportional to the intensity of the Bragg peak at that location.  Blue and purple vectors indicate the projections of the 6D reciprocal lattice basis vectors.  Cyan vectors indicate that a sum of three basis wavevectors yields a wavevector in the direction of one of another basis wavevectors, but scaled by $\phi$.  Red circles mark two rings of high intensity peaks that highlight the chiral nature of the pattern.  Purple circles are peaks with the same 6D indices as those shown in Fig.~\ref{fig:diffGolden} below.  In the generic case, these peaks are not related by a mirror symmetry.}
\label{fig:diffgeneric}
\end{figure}

Figure~\ref{fig:diffgeneric} shows a diffraction pattern for the generic case of Fig.~\ref{fig:keytiles}.  The peaks shown correspond to all 6D reciprocal lattice points $2\pi (k_1,k_2\ldots k_6)$ with $-4\leq k_n \leq 4$.  The elements of the reasoning outlined above are explained in more detail in the following paragraphs. 

\begin{figure}
\includegraphics[width=0.8\columnwidth]{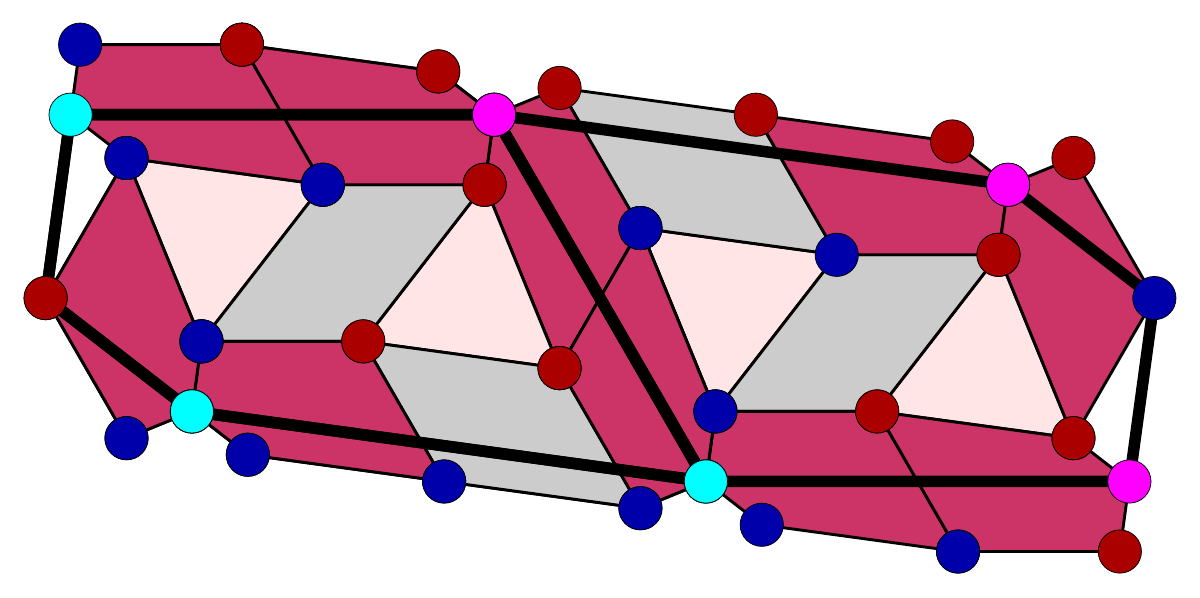}
\caption{Labeling of Key tile vertices by values of projection onto the $\Gamma$ subspace.  Colors represent pairs of components in the $\gv_1$ and $\gv_2$ directions:  magenta: $(0,1)/\sqrt{3}$; dark red $(0,0)$ dark blue: $(-1,0)/\sqrt{3}$; cyan: $(-1,-1)/\sqrt{3}$.}
\label{fig:Gverts}
\end{figure}

To see that the tiling vertices project onto only four distinct points in the $\Gamma$ subspace, first observe that every vertex in the tiling lies on at least one $F$ tile.  Figure~\ref{fig:Gverts} shows the deflation of a pair of $F$ tiles with vertices labeled according to their $\Gamma$ coordinates.  Moving along the direction $\ev_1$, $\ev_2$, or $\ev_3$ causes $g_1$ to increase by $1/\sqrt{3}$; moving along $\ev_4$, $\ev_5$, or $\ev_6$ causes $g_2$ to increase by $1/\sqrt{3}$.  Note also that vertices connected by a $c$ edge project to the same point in $\Gamma$.  Examination of the configurations in which a $P$ tile connects two $F$ tiles reveals that all $F$ tiles in orientations related by rotation by $\pm2\pi /3$ have vertices with one pattern of $(g_1,g_2)$ values, and all $F$ tiles rotated by $\pi/6$ from those three orientations have a single pattern.  The two patterns are displayed in Figure~\ref{fig:Gverts}, one $F$ tile having one magenta, three dark red, and one dark blue vertex and the other having one cyan, three dark blue, and one dark red vertex.

\begin{figure}
\includegraphics[width=\columnwidth]{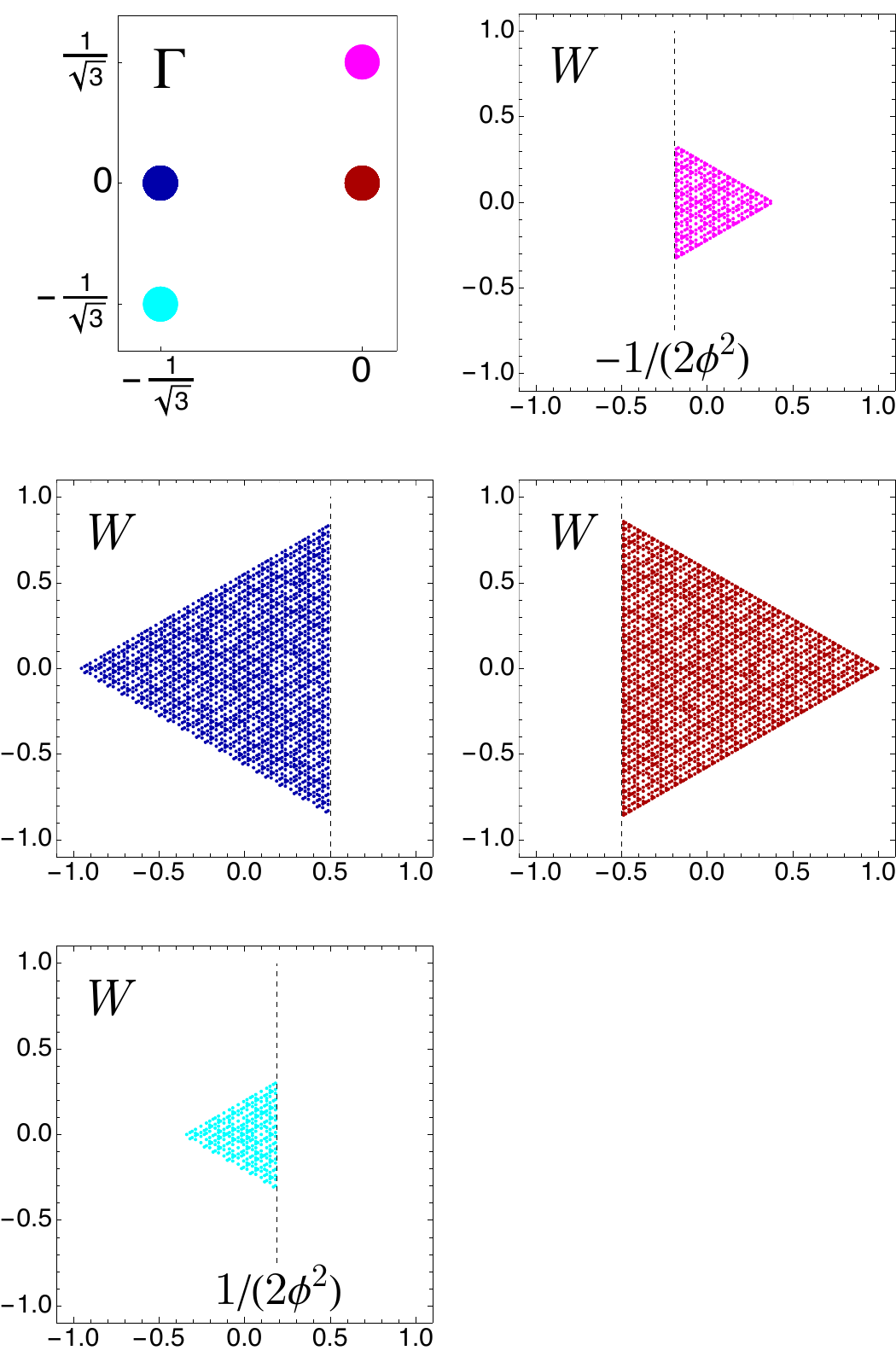}
\caption{Projection of the 6D tiling vertices onto the $\Gamma$ and $W$ spaces.  These images show the projections of 5658 vertices obtained by five deflations of a cluster of four tiles.  Colors of points in the four regions of $W$ correspond to the colors of the projections of those points in $\Gamma$.}
\label{fig:Window}
\end{figure}

An analytical derivation of the boundaries of the region containing the projection of the 6D tiling vertices onto the $W$ subspace is beyond our present scope.  We content ourselves here with an extrapolation from numerical data obtained by projecting a large finite portion of a tiling onto $W$.  Figure~\ref{fig:Window} shows the result.  It appears clear that the projection consists of four uniform density equilateral triangles.  The triangles associated with $\sqrt{3}(g_1,g_2) = (0,0)$ and $(-1,0)$ have areas $\phi^4$ larger than than those associated with $(0,1)$ and $(-1,-1)$  We can confirm that they have the same density by calculating the frequency ratio of different color vertices in a tiling decorated as in Fig.~\ref{fig:Gverts}.  Noting that the $(-1,-1)$ and $(0,1)$ vertices are the ones at the intersection of two $b$ edges in an $F$ tile, counting the numbers of different vertex types contributed by each deflated tile in Fig.~\ref{fig:keytiles}, and using the tile frequency rations from Eq.~\ref{eqn:ratios}, we find
\begin{align}
\frac{\#\ {\rm (-1,0)} + \#\ {\rm (0,0)}}{\#\ {\rm (0,1)} + \#\ {\rm (-1,-1)}}  & = \frac{6 \rho_H + 3\rho_T + 6\rho_P + \frac{19}{3}\rho_F}{\rho_T + \frac{4}{3}\rho_P + \frac{4}{3}\rho_F} \nonumber \\
& = \phi^4,
\end{align}
confirming that the uniform density hypothesis is plausible.

Let $\xv$ be a 6D lattice point and let $\xv_t$, $\xv_w$, and $\xv_g$ be its projections onto the tiling plane, $W$, and $\Gamma$, respectively. We want to compute the structure factor for a set of unit masses placed at the vertices of the Key tiling:
\begin{equation}\label{eqn:Sofk1}
S(\kv_t) = \sum_{\xv_t} e^{i \kv_t\cdot\xv_t}.
\end{equation}
The Fourier transform of the hypercubic lattice consists of delta-function peaks of equal amplitude at all points $\kv$ of a hypercubic reciprocal lattice for which $e^{i \kv\cdot\xv} = 1$ for all lattice vectors $\xv$.  Each $\kv_t$ in the tiling plane is specified by the 6 integers defining its lift into the 6D hypercubic reciprocal lattice.  We then define 
\begin{equation}
(\kv_t,\kv_w,\kv_g) = (\Pm^{-1})^T \kv
\end{equation}
so that
\begin{equation}
 \kv_t \cdot\xv_t+ \kv_w\cdot\xv_w + \kv_g\cdot\xv_g = \kv \Pm^{-1}\Pm\xv = \kv\cdot\xv.
\end{equation}
A straightforward calculation (performed analytically using Mathematica~\cite{Mathematica}) reveals that 
\begin{equation}
(\Pm^{-1})^T (0,0,2,0,1,0) = (\Pm^{-1})^T (0,0,0,0,0,\phi),
\end{equation}
indicating that an integer linear combination of reciprocal space basis wavevectors $\kv_3$ and $\kv_5$ yields a wavevector in the direction of $\kv_6$ but scaled by $\phi$, as illustrated by the cyan arrows in Fig.~\ref{fig:diffgeneric}.  The relation holds for any values of $a$, $b$, and $\theta$, indicating that the ratio characterizing the incommensurate wavelengths present in the tiling remains locked to the golden mean as the Key tile shapes are varied continuously.

We can now write 
\begin{align}
S(\kv_t) & = \sum_{\xv\in L} e^{i \kv\cdot\xv} e^{-i \kv_w\cdot\xv_w}e^{-i \kv_g\cdot\xv_g} \label{eqn:Sofk} \\
\ & = \sum_{(g_1,g_2)} \left(e^{-i \kv_g\cdot\xv_{(g_1,g_2)}}\int_{\Delta(g_1,g_2)} \!\!\!\!\!\!\!\!\!\!\!\!\!\!e^{-i \kv_w\cdot\xv_w } d\xv_w \right),
\label{eqn:Sofkperp}
\end{align}
where the sum in Eq.~(\ref{eqn:Sofk}) runs over all of the hypercubic lattice points in the lift, $L$, of the tiling; the sum in Eq.~(\ref{eqn:Sofkperp}) runs over the four $(g_1,g_2)$ pairs defined above; and $\Delta(g_1,g_2)$ is the uniform density triangle in $W$ containing the points associated with the given $(g_1,g_2)$ pair.

The Fourier transform of the triangular region in Fig.~\ref{fig:Window} corresponding to $\xv_g = (0,1)$ (dark red), is obtained by straightforward integration, yielding
\begin{widetext}
\begin{equation}
{\cal F}_{(-1,0)}(\kv_w) = \frac{
\sqrt{3} k_{w,y}e^{-i k_{w,x}}  + 
   \left( 3 i k_{w,x} \sin \left(\tfrac{\sqrt{3}}{2} k_{w,y}\right) - \sqrt{3} k_{w,y} \cos \left(\tfrac{\sqrt{3}}{2} k_{w,y}\right)\right) e^{i k_{w,x}/2}
   }{
   \pi k_{w,y} \left(k_{w,y}^2-3 k_{w,x}^2\right)}.
\end{equation}
\end{widetext}
The transforms of the other windows are 
\begin{align}
{\cal F}_{(0,0)}(k_{w,x},k_{w,y}) & = {\cal F}_{(-1,0)}(-k_{w,x},k_{w,y}); \\
{\cal F}_{(-1,-1)}(k_{w,x},k_{w,y}) & = {\cal F}_{(-1,0)}(k_{w,x}/\phi^2,  k_{w,y}/\phi^2);  \nonumber \\
{\cal F}_{(0,1)}(k_{w,x},k_{w,y}) & = {\cal F}_{(-1,0)}(- k_{w,x}/\phi^2, k_{w,y}/\phi^2). \nonumber
\end{align}
We then have
\begin{align}
S(\kv_t) =  {\cal F}_{(0,0)}(\kv_w) & + e^{i \kv_g\cdot(-\gv_1)} {\cal F}_{(-1,0)}(\kv_w)   \\ 
 & + e^{i \kv_g \cdot\gv_2}{\cal F}_{(0,1)}(\kv_w) \nonumber \\
 & + e^{i \kv_g \cdot(-\gv_1-\gv_2)}{\cal F}_{(-1,-1)}(\kv_w), \nonumber
\end{align}
where 
\begin{align} 
\gv_1 & \equiv (1,1,1,0,0,0)/\sqrt{3}, \\
\gv_2 & \equiv (0,0,0,1,1,1)/sqrt{3}.
\end{align}  

\begin{figure}
\includegraphics[width=\columnwidth]{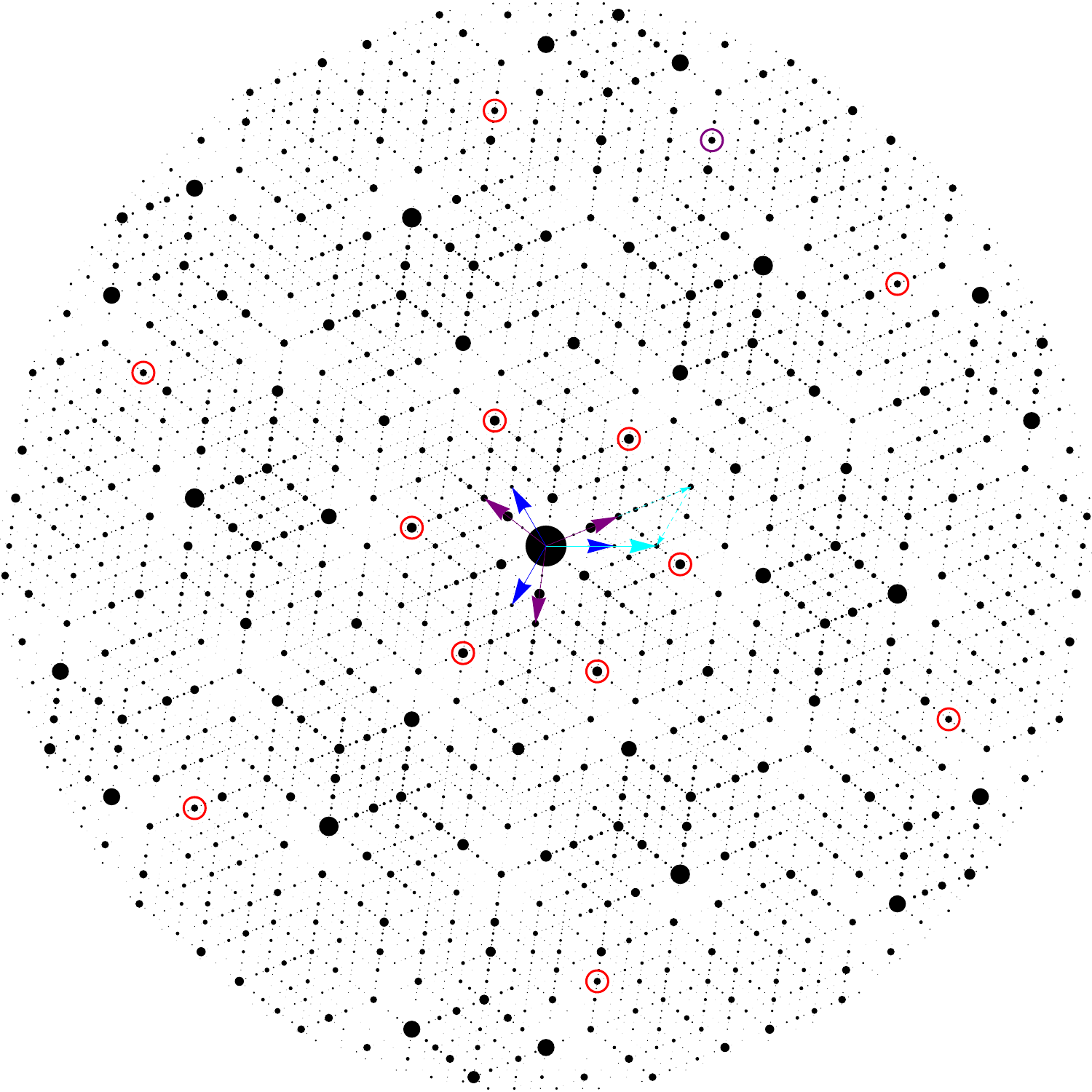}
\caption{The diffraction pattern of unit masses placed at the vertices of the Golden Key tiling.  Blue and purple arrows indicate the projections of the 6D reciprocal lattice basis vectors.  Cyan vectors indicate that a sum of 3 basis wavevectors yields a wavevector in the direction of one of the basis vectors and scaled by $\phi$.  Red and purple circles mark the peaks with the same indices as their counterparts in Fig.~\ref{fig:diffgeneric}.  Here their wavevectors and intensities are related by a reflection symmetry corresponding to the mirror symmetry of the set of tiling vertices.}
\label{fig:diffGolden}
\end{figure}

\begin{figure*}
\includegraphics[width=\textwidth]{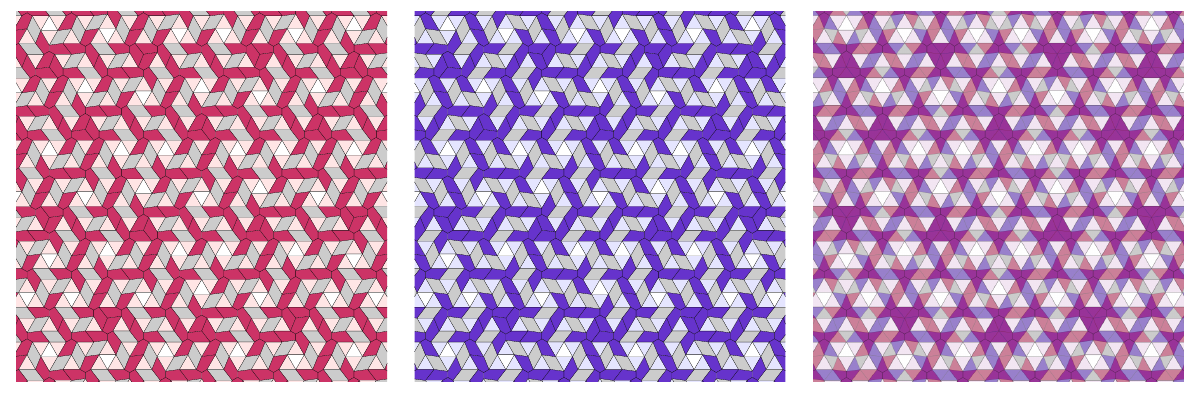}
\caption{The Golden Key tiling, its mirror image, and an overlay of the two.  Note that the two tilings have exactly the same set of vertices, which is not a chiral set.  In each tiling, the mirror symmetry is broken by the choice of which vertices are connected to form the $P$ and $F$ tiles.}
\label{fig:GoldenTM}
\end{figure*}

The calculated diffraction pattern for a generic Key tiling is shown above in Fig.~\ref{fig:diffgeneric}, and it displays the chirality that is apparent in the tiling itself.
Figure~\ref{fig:diffGolden} shows the calculated diffraction pattern for the Golden Key tiling.  Somewhat surprisingly, the diffraction pattern displays mirror symmetries even though the tiling is clearly chiral.  
This is because the pattern of vertices is actually not chiral; the grouping of vertices to form tiles creates a chiral pattern, but, rather remarkably, one can connect the same vertices with a mirror image set of tiles, as shown in Fig.~\ref{fig:GoldenTM}.

\begin{figure}
\includegraphics[width=\columnwidth]{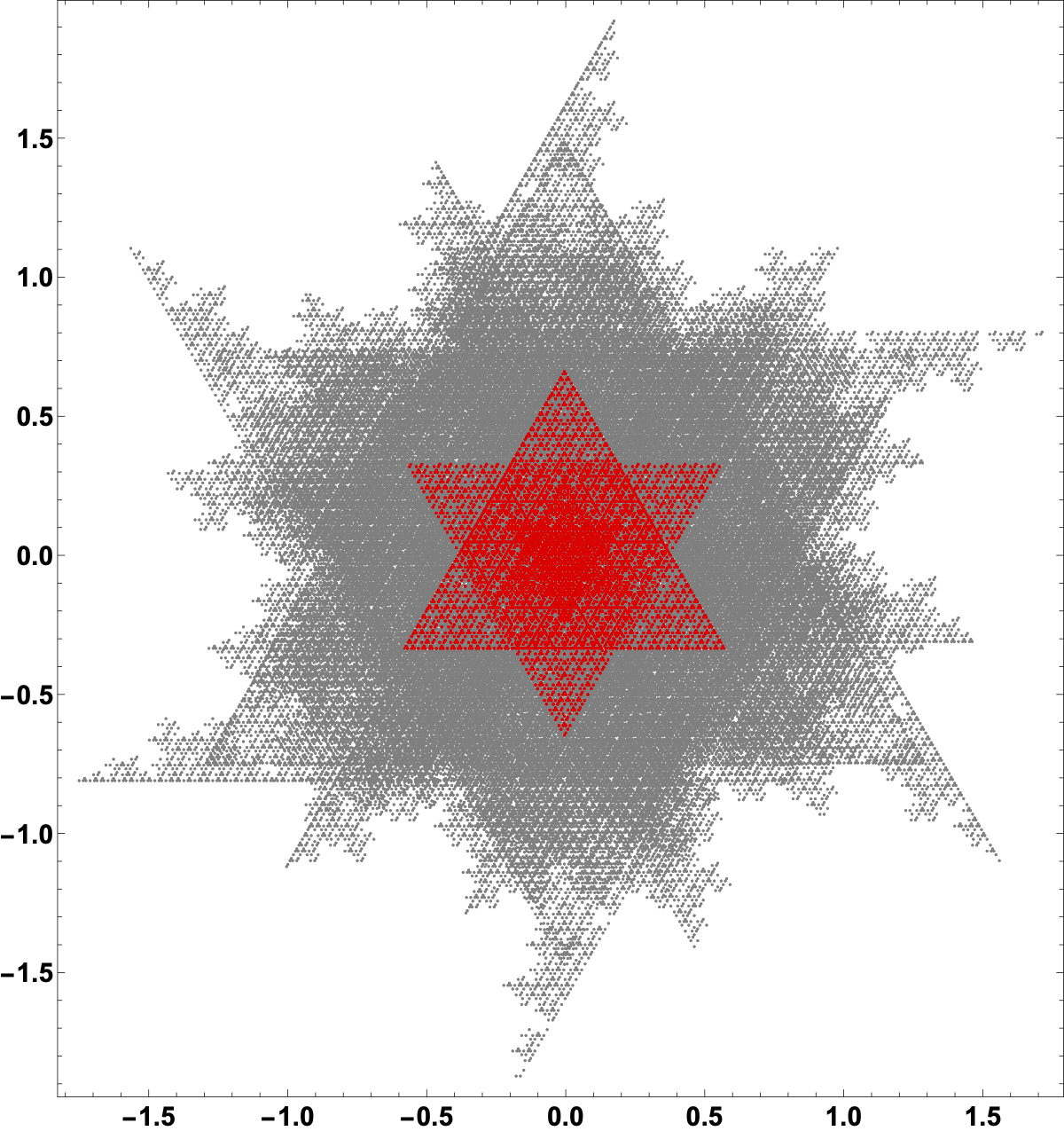}
\caption{The projection of Hat tiling vertices onto $W$.  Gray points are Hat vertices; red points are the subset that are vertices of the associated Key tiling.  Planes corresponding to different points in $\Gamma$ are superimposed here.  Note that the sizes and orientations of the triangular regions here differ from above because the basis vectors of the 6D lattice correspond to Hat edges rather than Key tile edges.}
\label{fig:hatW}
\end{figure}

For completeness, we show in Fig.~\ref{fig:hatW} the projection of the vertices of a Hat tiling onto $W$.  Here we have defined the lift to the hypercubic lattice using the edges of the hats, and we have collapsed the regions correspond to different points in $\Gamma$ onto a single plane.  The vertices of the Key tiles, shown in red, are a subset of the points in this lift of the Hat tiles.  (See Figs.~\ref{fig:FedgesHat} and~\ref{fig:HatTilings}.) 
Note that the acceptance window is substantially more complex than that of the Key tiling and itself has a chiral structure.  One thus expects the diffraction intensities of the Hat tiling to display a more exaggerated chiral structure than that of the corresponding Key tiling.
This and other decorations of the Key tiles with point masses or continuous densities will yield a diffraction pattern with Bragg peaks at the same wavevectors but with intensities modified by form factors associated with the different types of vertices in the tiling.  (See the remark below concerning the almost unique determination of tile orientations from Key tile vertex locations.)

\section{Phason rearrangements: worms and snakes}\label{phasons}
In the theory of Penrose tilings, the so-called cartwheel tiling plays an important role.~\cite{gardner1977}  It is an infinite tiling that is 10-fold symmetric except for the tiles covering 10 infinite rays emanating from its center.  These tiles make up ``worms'' that are internally rearranged (or ``flipped'') under infinitesimal phason shifts~\cite{socolar1986PPD}.  Here we construct the analogue of the Penrose cartwheel and use it to reveal the phason flips in the Key and Hat tilings.
 
\begin{figure*}
\includegraphics[width=\textwidth]{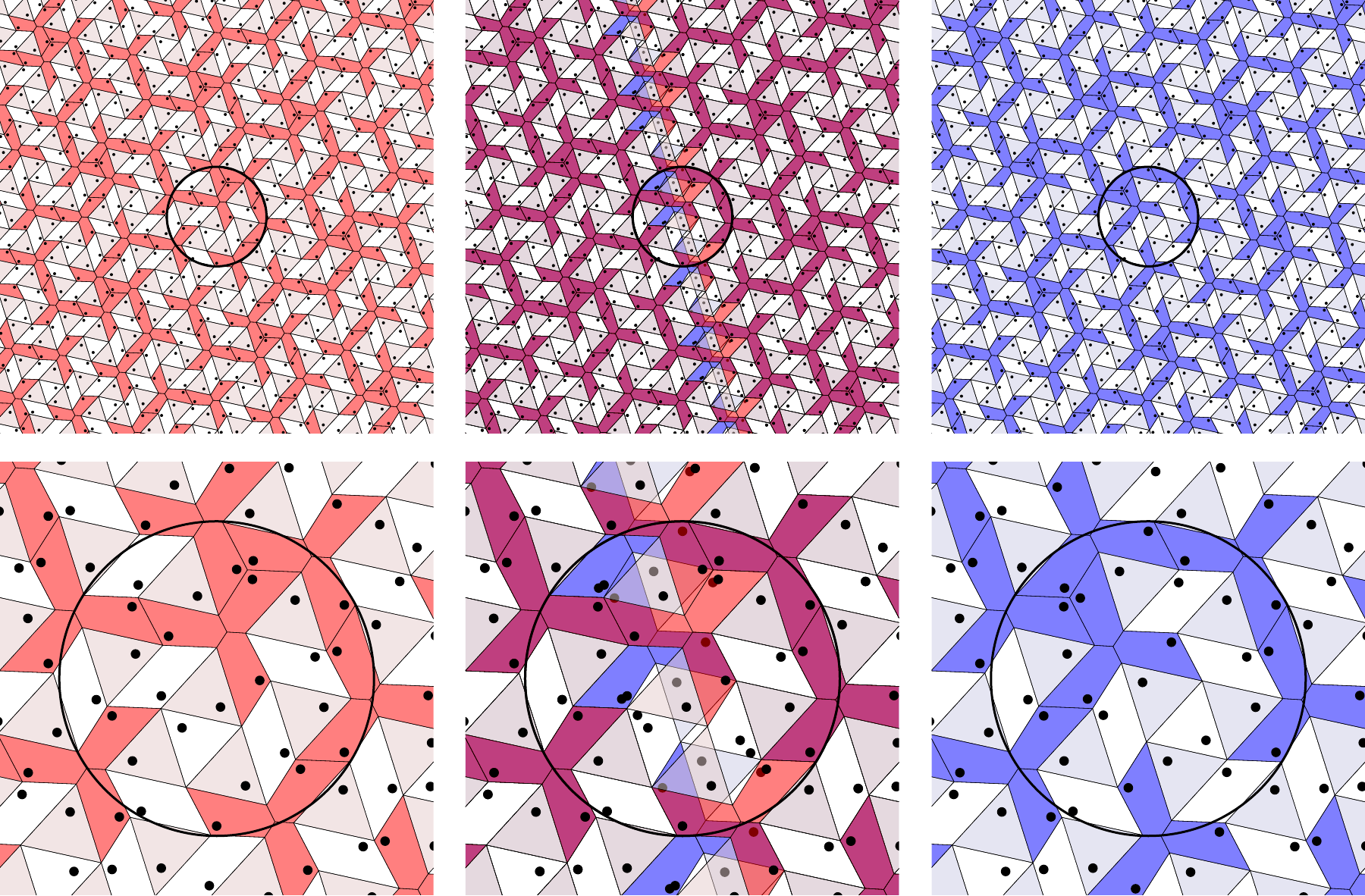}
\caption{A portion of a generic Key tiling illustrating one type of tile rearrangement associated with an infinitesimal phason shift.  Dots indicate the orientations of tiles, which provide necessary information for constructing the Hat decoration.  Left: a tiling formed from multiple deflations of the configuration contained within the encircled region, with $F$ tiles assigned a semi-transparent red color.  Right: a tiling obtained in the same way, but where the central region is rotated with respect to its counterpart on the left and differs from it in the orientations of some of the tiles, with $F$ tiles assigned a semi-transparent blue color.  Center:  an overlay of the flanking panels.  Coinciding red and blue regions become purple, revealing that the tilings differ only by the rearrangement of tiles along an infinite worm.  Bottom panels are scaled versions of the corresponding top panels.}
\label{fig:Cw}
\end{figure*}

\begin{figure}
\includegraphics[width=1.0\columnwidth]{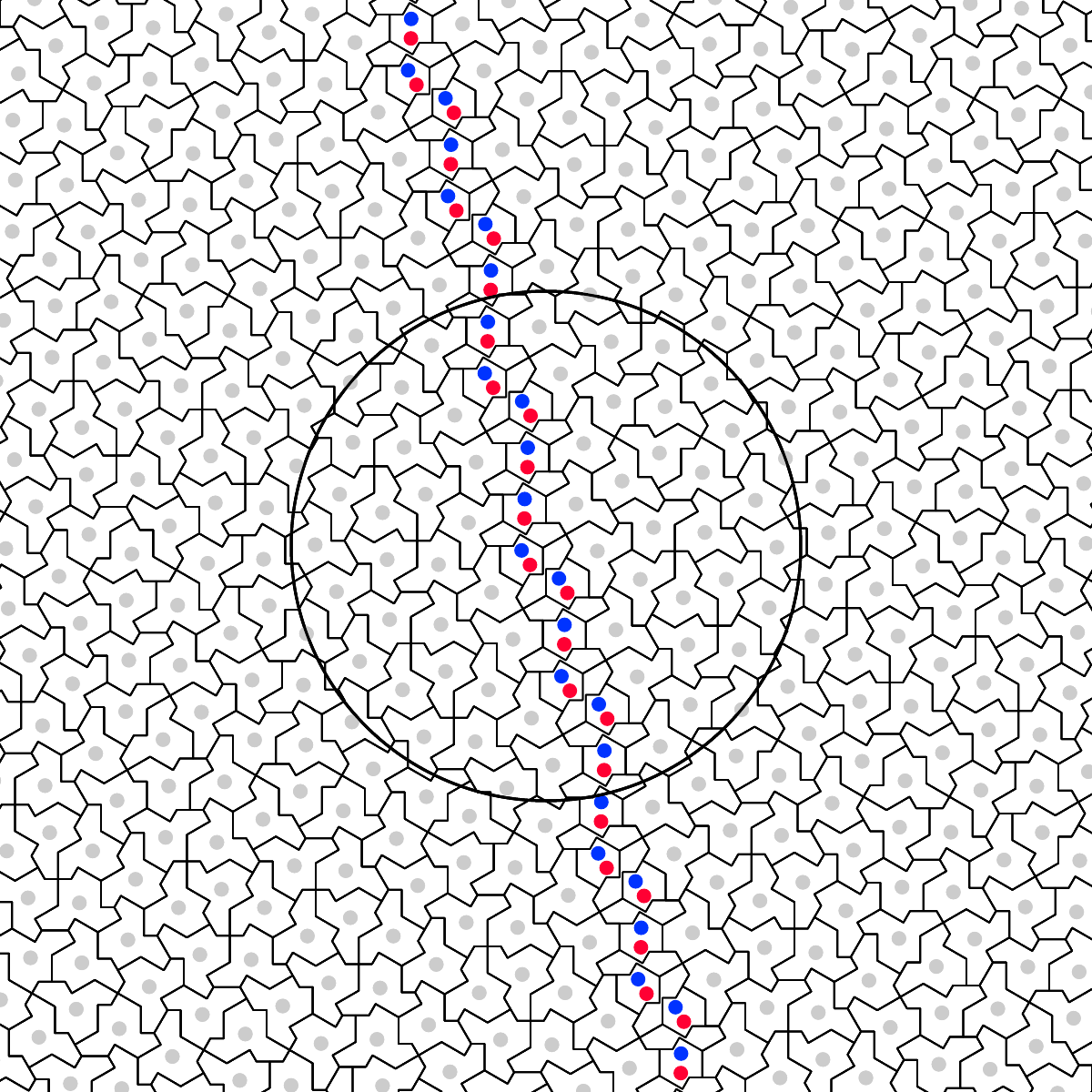}
\caption{A portion of a Hat tiling illustrating a worm of tile rearrangements associated with an infinitesimal phason shift.  Red and blue dots mark the centers of hats decorating the two different defect-free tilings of Fig.~\ref{fig:Cw} with lengths scaled as indicated by the corresponding central circles.  Light gray dots indicate tiles where the two tilings coincide. The tilings differ only by Hat rearrangements along an infinite worm passing through the central circle. } 
\label{fig:HatCw}
\end{figure}

Figure~\ref{fig:Cw} shows two tilings that differ by the flip of a single infinite worm.  
The red panel shows a cartwheel tiling where six semi-infinite ``worms'' intersect at the central disk.  The blue panel shows a configuration in which one of those worms has been flipped, corresponding to the infinitesimal shift in the location of the window within the $W$ subspace.  The central panel shows a superposition of the two, with purple indicating coincidence of the two patterns, showing that the two differ by the reconfiguration of a single worm.  The surprise comes when one examines the orientations of the tiles more carefully.  If this were an exact analogue of the Penrose case, the pattern would have complete hexagonal symmetry when the three infinite worms and central disk are removed, and indeed it is true that the tile pattern exhibits this symmetry.  However, the tile orientations, as indicated by the small dots, do not respect this symmetry.  This can be seen most easily by inspection of the six triangular $H$ tiles that border the outside of the central circle in Fig.~\ref{fig:Cw}, where only one of them has a dot at the vertex farthest from the circle.  Because these dots determine how the Hat decorations are placed, the Hat tiling derived from this Key tiling does not exhibit the expected symmetry.  The fact that the Key tiling with orientation dots removed can exhibit the hexagonal symmetry has it origin in the symmetry of the deflation rule for the $P$ tiles.  As shown in Figs.~\ref{fig:deflation} and~\ref{fig:HatTilings}, the deflated tile configuration is symmetric under rotation by $\pi$, but the Hat decoration of the $P$ tile is not.   Figure~\ref{fig:HatCw} shows the superposition of the Hat tilings derived from the two cartwheel tilings of Fig.~\ref{fig:Cw}, revealing the structure of the worm in the Hat tiling.  Each Hat is marked with a disk at its center, with red indicating a Hat from one tiling, blue indicating a Hat from the other, and gray indicating coincidence of the two.  These observations may point the way to development of a local growth algorithm for constructing an infinite Hat tiling from a central defective seed similar to the Penrose tiling growth algorithm~\cite{onoda1988,socolar1991}. 

\begin{figure}
\includegraphics[width=1.0\columnwidth]{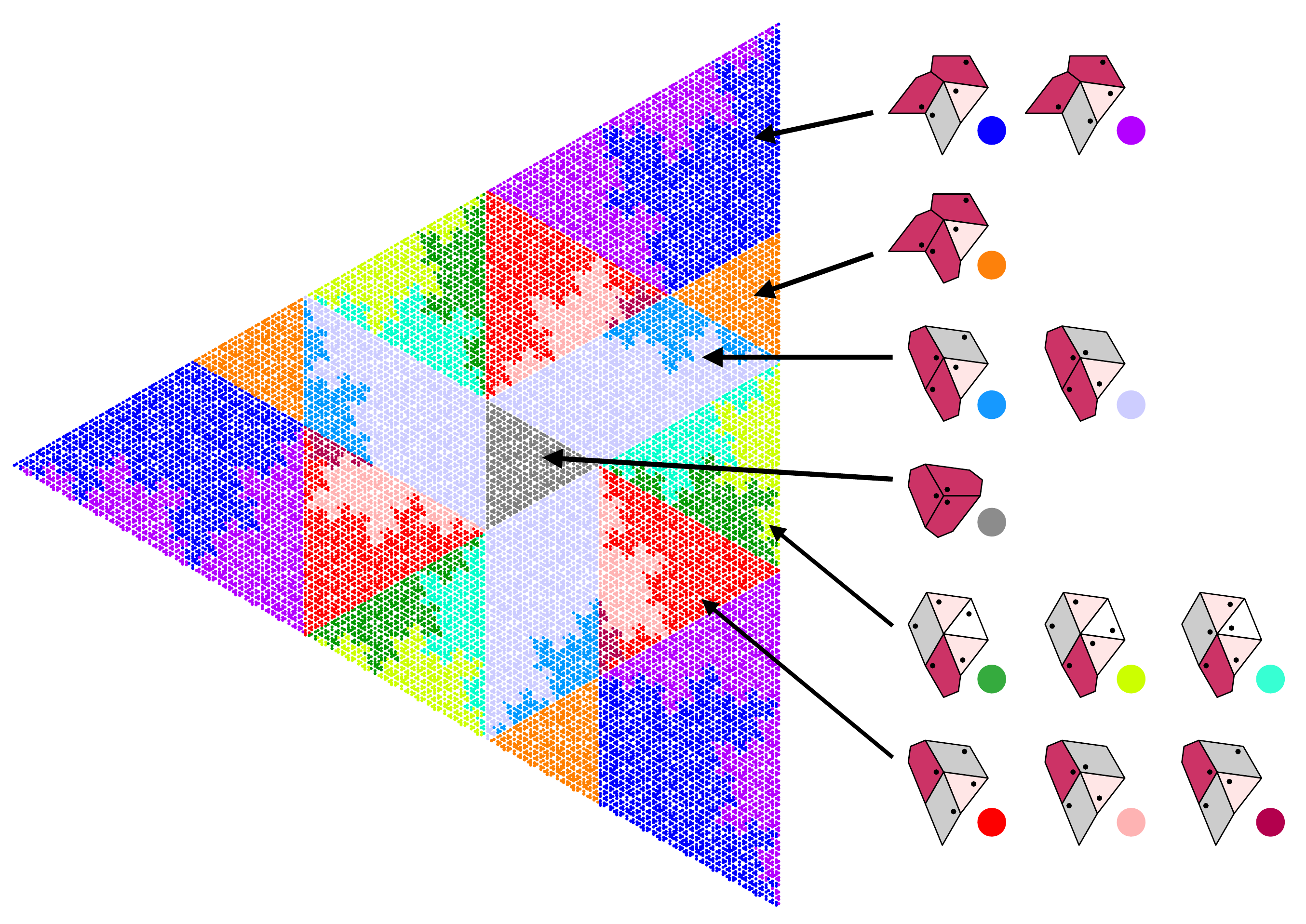}
\caption{One sheet of the 4D window showing the regions corresponding to different vertex configurations.  The color of a dot indicates the vertex configuration shown on the right.  Regions related by rotation correspond to vertex configurations related by rotation.  Note that vertices with identical surrounding Key tile configurations but different tile orientations are separated by fractal boundaries.  The different orientations give rise to different patterns of hats, as shown in  Fig.~\ref{fig:HatSnake} below.} 
\label{fig:HatVertexTypes}
\end{figure}

\begin{figure}
\includegraphics[width=1.0\columnwidth]{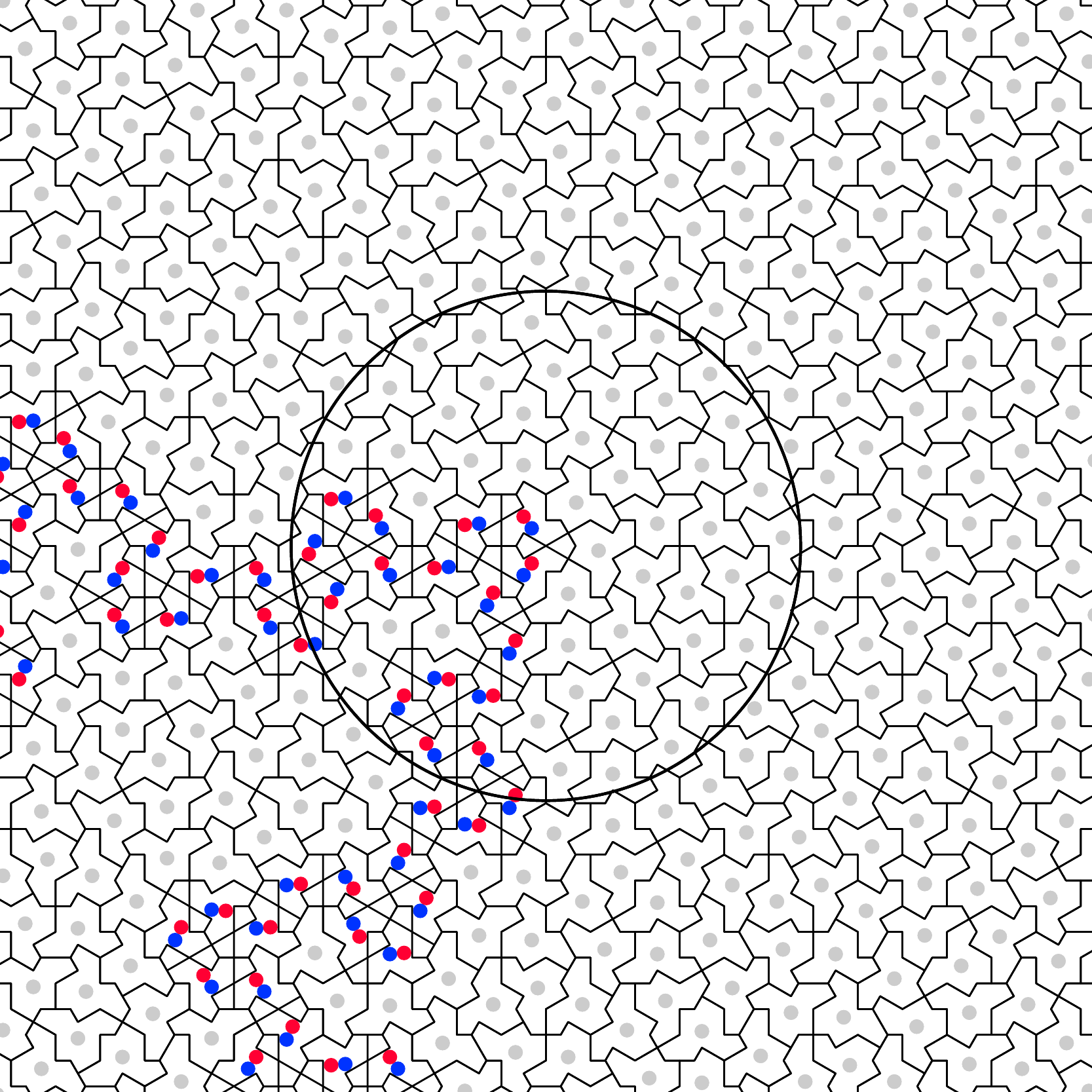}
\caption{A portion of a Hat tiling illustrating a snake of tile rearrangements associated with an infinitesimal phason shift.  Red and blue dots mark the centers of hats decorating two different defect-free tilings, one of which is the red tiling of Fig.~\ref{fig:Cw}, with lengths scaled as indicated by the corresponding central circles.  Light gray dots indicate tiles where the two tilings coincide. The tilings differ only by Hat rearrangements along an infinite snake passing through the central circle. } 
\label{fig:HatSnake}
\end{figure}

The phason rearrangements described above are comprised of Key tiles sharing a set of vertices that project onto the straight-line boundaries of the projection window.  There is, however, a different type of phason rearrangement comprised of tiles that share vertices lying on the fractal boundaries that separate different types of vertices in the interior of the window.  Figure~\ref{fig:HatVertexTypes} shows one sheet of the window in which vertices have been coded according to the configuration of tiles surrounding them.  
The fractal boundaries separate vertices with identical tile shapes but different tile orientations, indicated by the black dots.  
(These fractal boundaries are the origins of the fractal boundary of the window for the Hat vertices shown in Fig.~\ref{fig:hatW}, but the detailed relationship is beyond our present scope.)
A translation of the window that causes points to cross such boundaries results in a rearrangement of Hat tiles along a path that we refer to as a ``snake.''  Figure~\ref{fig:HatSnake} 
shows a superposition of two cartwheel tilings that differ only by the rearrangement of tiles along a snake.  The two Key tilings are identical except for the locations of the orientation dots along the snake.

We note that the $H$, $T$, and $P$ tiles in the Key tiling form a network in which every branch may be a segment of a snake and there are no closed loops.  By inspection, we see that the orientation of the $H$ tile at the tip of a branch is fixed, so a path can only be flipped if it is infinite in both directions; i.e., only the infinite snake can be flipped.  This observation implies that the orientation of the Key tiles is completely determined by the locations of the tile vertices with the possible exception of the tiles on an infinite snake.  This in turn confirms that the diffraction pattern calculated from the Key tile vertex locations contains all of the information required to construct the Hat tiling itself, which is consistent with the rigorous result of Baake et al.\ showing that the diffraction from a Key tiling in which all (oriented) tiles of the same type are decorated with the same (oriented) mass distribution consists purely of Bragg peaks at the same locations as the pattern calculated above ~\cite{baake2023}.

\section{Remarks}\label{remarks}
In an effort to understand the structure of the Hat tilings, we have discovered a broader, two-continuous-parameter class of quasicrystalline tilings with $C_6$ symmetry.  These ``Key tilings'' are all constructed by projection of the same set of 6D hypercubic lattice points onto the tiling plane; i.e., those that project onto a certain window in the 4D subspace spanned by $W$ and $\Gamma$.  The difference between tilings consisting of different shape Key tiles is the orientation of the tiling plane with respect to the hypercubic lattice.  The tiling plane must be orthogonal to $\Gamma$ but is not constrained by symmetry to be orthogonal to $W$.  A continuum of possible orientations yield geometrically (but not topologically) distinct tilings, all of which inherit a substitution (or inflation) operation from the substitution procedure defined on the 6D lattice.  In general, the substitution operation does not preserve the shapes of the Key tiles, and repeated deflations eventually produce self-intersecting tile perimeters.  In the special case of the Golden Key tiles, the tile shapes are preserved, allowing for infinite iteration of the deflation operation.  In this case (and only in this case) the set of tiling vertices exhibits mirror symmetry.

As noted by Smith et al., the Hat tilings are simple decorations of a one-parameter family of metatiles or, equivalently, of the Key tiles.  This family is embedded in a 2-parameter class.  Other members of the 2-parameter class admit decorations consisting of two Hat tiles that are not mirror images of each other but yield tilings combinatorially equivalent to the Hat tilings.  Still others, including the Golden Key tilings, do not admit such decorations.  Whether they may admit decorations by some other shape that reduces the number of tile types from four to two — or perhaps even one — is an open question.

The view from six dimensions reveals that these tilings are all quasicrystalline, having diffraction patterns consisting of a dense set of Bragg peaks.  The set includes incommensurate wavevectors with wavelengths form $(n + m\phi)k_0$, where $\phi$ is the golden mean, $k_0$ is the wavenumber of one of the basis vectors, and $n$ and $m$ are integers.  Although the hexagonal point group symmetry does not necessarily force a particular incommensurate ratio, the ratio $\phi$ remains fixed as the parameters of the Key tiles are varied continuously.  This locking of the incommensurability justifies the ``quasicrystalline'' label, distinguishing these hexagonal structures from those exhibiting incommensurate density waves with continuously variable wavelength ratios.  We note also that this structure is qualitatively different from the limit-periodic structure of the hexagonal Taylor-Socolar monotile tiling, which has Bragg diffraction peaks at wavelengths of the form $2^{-n}k_0$ for arbitrarily large $n$.~\cite{marcoux2014}

In the Penrose case, the 10-fold symmetry of the system forces the incommensurate wavelengths to be related by the golden mean.  There is no obvious reason, however, for the ratio $\phi$ to be uniquely selected for structures with hexagonal symmetry.  Analogous constructions may exist in which the subspace $W$ is oriented so as to produce different locked incommensurate ratios.  One may also wonder whether an analogous construction based on a 4D hypercubic lattice might produce a class of golden mean tilings with square symmetry — and possibly one that admits another monotile decoration.

The projection construction provides a window into many properties of the Hat tilings.  All of the projection techniques that have been brought to bear in the analysis of Penrose tilings and other quasicrystals can be adapted to the Key tilings and, by extension, to the Hat tilings.  These include methods for analyzing the empires of finite patches of the tiling~\cite{gardner1977,fang2017}, the tile configurations associated with phason defects and dislocations~\cite{socolar1986PPD}, and the possible development of local growth algorithms.

\noindent {\it Acknowledgement:}  The author thanks Paul Steinhardt for helpful comments on drafts of the manuscript and two anonymous referees for comments that significantly improved the originally submitted paper.

\bibliography{KeyTiling}

\end{document}